# PATTERN BASED ADAPTIVE ARCHITECTURE FOR INTERNET BANKING

[1]A. Meiappane, [2]V. Prasanna Venkatesan, [3] V. Jegatheeswari
[4]B. Kalpana and [5]U. Sarumathy

[1]Research Scholar, [2]Associate Professor, Pondicherry University, Pondicherry
[3, 4, 5]Student, Department of IT, Sri Manakula Vinayagar Engineering College, Pondicherry
E-mails: [1]auromei@yahoo.com, [2]prasanna_v@yahoo.com, [4]kalpuji@gmail.com

*Abstract:* Pattern plays a vital role in software architecture and it is a general reusable solution to commonly occurring problem. Software architecture of a system is the set of structures needed to reason about the system, which comprise software elements, relations among them, and properties of both. Patterns can be implemented at run-time; they identify key resource constraints and best practices. Architecture Pattern expresses a fundamental structural organization or schema for software systems. Patterns in software architecture, offer the promise of helping the architect to identify combinations of Architecture or Solution Building Blocks that have been proven to deliver effective solutions. In Internet banking, we analyzed some attributes such as reliability, security, availability, load balancing and so on. The use of patterns, which is of a reusable component, is a good tool to help designers build load balancing systems. In this paper we are going to propose pattern based adaptive architecture for internet banking system and so the above attributes can be improved by the usage of patterns.

## 1. INTRODUCTION

### 1.1. Software Architecture

It describes the structure of the components of a program or system and the relationships between them. Patterns in software architecture are a useful method to describe a good solution to a recurring problem. The software architecture of a program o the computing system is the structure of the system, which comprises software components, the externally visible properties of those components, and the relationship between them. Architecture supports the goal of the system and respects their constraints. Architecture identifies the critical issues and makes explicit the design choices made. Finally architecture prescribes the rules that must be followed. Architecture of the software system is the one which drives the implementation. It shifts the focus from the lines of code to the software components and their interconnection. [6]

### 1.2. Pattern

It is defined as an idea that has been useful in one practical context and will probably be useful in others. Patterns are considered to be a way of putting building blocks into context. For example to describe a reusable solution to a problem. Building blocks are what you use: patterns can tell you how you use them, when, why and what trade-offs you have to make in doing so [8].

Patterns offer the promise of helping the architect identify combinations of architectural and/or solution building blocks that have been proven to deliver effective solution in the past, and may provide the basis for effective solutions in the future.

Software and building architects have many similar issues to address, and so it as natural for software architects to take interest in patterns as architectural tools [3].

The use of patterns has become increasingly popular for many domains, including those within the computing field. Developing patterns in a given domain offers something interesting and unique to practitioners and researches. For practitioners, patterns offer practical and applied knowledge by providing high level solutions to classes of problems that can be converted into specific best practices. For researches, patterns can provide a method to synthesize and capture knowledge in a given domains as well as highlight areas for future research.

A particular recurring design problem that arises in specific design contexts, and presents a well proven generic scheme for its solution. The solution scheme is specified by describing its constituent components, their responsibilities and relationships, and the ways in which they collaborate. A re-usable solution to a recurring problem, tried, tested and can be adapted, personalized for the problem domain. Three categories of patterns

- Architectural patterns
- Design patterns
- Idioms

### 1.2.1. Architectural Patterns

A high-level structure for software systems and contains a set of predefined sub-systems which also defines the responsibilities of each sub-system, details the relationships between sub-systems. Also similar to 'conceptual patterns' which cover the application domain.

An architectural pattern expresses a fundamental structural organization or schema for software systems. It provides a set of predefined subsystems specifies their responsibilities and includes rules and guidelines for organizing the relationships between them.

Architectural patterns are templates for concrete software architectures.

### 1.2.2. Design Patterns

It is Mid-level construct and it is implementation-independent, designed for 'micro-architectures'– somewhere between sub-system and individual components, design patterns are medium-level strategies that are concerned with the structure and behavior of entities, and their relationships. Design patterns do not influence the overall system structure, but instead define micro architectures of subsystems and components.

### 1.2.3. Idiom Patterns

Earliest form of software pattern, comparatively low-level and gives a guide for implementing the components and relationships of the pattern. Considers the pattern at a programming language level (Describes the pattern using the constructs of the specific language)

## 2. FRAMEWORK

A framework is an integrated set of software artifacts (such as classes, objects, and components) that collaborate to provide a reusable architecture for a family of related applications .It is used in a wide range of different domains, such as telecommunications, avionics, manufacturing, and financial services. Framework is a set of co-operating classes that make up a reusable design for a specific class of software.

Many approaches have addressed adaption using framework. However, they are not based on any systematic approach. Due to adhoc changes to the system, the architecture might deteriorate. This paper proposes an approach for dynamic self adaption which integrates the architectural pattern with the adaption oriented framework that enables the system to adapt according to some predefined and proven structure [11]

For adaptive systems to be built, frameworks have been developed.

### 2.1. Develop a New Framework

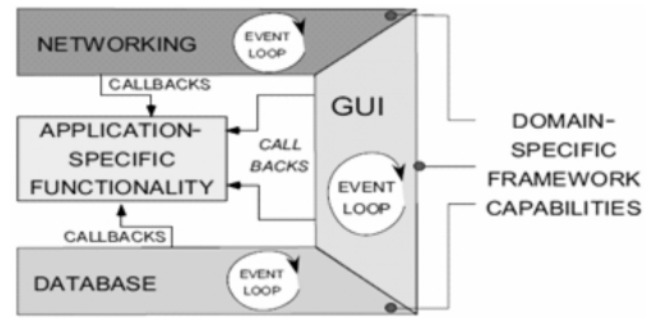

**Figure 1:** Relationships between Framework Artifacts

Figure 1 shows a key challenge of designing frameworks is to decompose the framework's capabilities into a set of reusable classes, while simultaneously anticipating future uses and changes [10].

Some specific issues that should be addressed when developing a new framework include determining: Which classes should be fixed, thus defining the stable shape and usage characteristics of the framework? If key interfaces in a framework aren't stable, it may be hard for users to understand and apply the framework effectively and efficiently because there will be too many degrees of freedom.

Which classes should be extensible, *e.g.*, by sub classing or template instantiation, to support adaptation necessary to use the framework for new applications. If a framework can't be extended, then users can't customize it for their needs, which makes it hard to accommodate a diverse set of applications and use cases that were not foreseen during the framework's initial design.

Determining the right protocols for startup and shutdown sequences of operations. If the application developers cannot pick and choose the initialization and termination sequences of framework operations, the lifetimes of the application and framework can get coupled in complex ways, which can reduce flexibility significantly.

Self adaption is in the form of programming language features embedded in the code, thus prohibiting reusability and modification.

A Self adaptive approach that integrates Monitoring, analyzing and actuation functionalities have the potential to accommodate to a dynamically changing environment.

### 2.1. Framework Advantages

The Framework has many advantages:

1. **Extensible** to support successions of quick updates and additions to address new requirements and take advantage of emerging markets,
2. **Flexible** to support a growing range of multimedia data types, traffic flows, and end-to-end quality of service(QoS) requirements,
3. **Portable** to reduce the effort required to support applications on heterogeneous OS platforms and compilers
4. **Reliable** to ensure that applications are robust and tolerant to faults
5. **Scalable** to enable applications to handle larger numbers of clients simultaneously
6. **Affordable** to ensure that the total ownership costs of software acquisition and evolution are not prohibitively high. [10]

### 2.2. Adaptive Framework

Adaptive Framework is a dynamic framework which monitors the running system and updates the architectural properties based on the changes in the system. An adaptive framework which has three key elements namely, monitoring, decision making and re-configuring. This framework works with the help of an external monitoring and control mechanism.

An advantage of any adaptation based framework is that it provides a lot of reusable code which enables the developers to build applications more quickly [1].

### 2.3. Adaptive Framework Benefits

- **Speed** of implementation with repeatable architectural patterns and accelerators
- **Flexibility** to progressively transform to a simplified architecture one project at a time
- **Choice** of how to get started and who to partner with for business capabilities
- **Cost Reduction** through re-use of services and assets and through faster implementation
- **Alignment** of business and IT priorities for more effective results from solution implementation[11]

### 3. INTERNET BANKING

Online banking (or Internet banking) allows customers to conduct financial transactions on a secure website operated by their retail or virtual bank, credit union or building society. The finance services, associated with Internet, are being promoted due to the wide spread use of Internet. Internet banking that is also known as online banking is one of the emerging services.

The Internet is rapidly turning out to be a tool of worldwide communication. The increasing use of Internet earlier promoted producers and entrepreneurs to sell their products online. It has also become an important source of information and knowledge. Due to this, many banking and finance organizations have come up with the idea of Internet banking or online banking. Internet banking can be defined as a facility provided by banking and financial institutions that enable the user to execute bank related transactions through Internet. The biggest advantage of Internet banking is that people can expend the services sitting at home, to transact business. Due to which, the account holder does not have to personally visit the bank. With the help of Internet banking many transactions can be executed by the account holder. When small transactions like balance inquiry, record of recent transaction, etc. are to be processed, the Internet banking facility proves to be very handy. The concept of Internet banking has thus become a revolution in the field of banking and finance. Many banking organizations had already started creating data ware housing facilities to ease their working staffs.

Internet banking is a system of banking that enables customers to perform various financial transactions on a secure website via the Internet. There are many banks and credit union that operate websites for internet banking. Internet Banking is basically conducted via a personal computer connected to Internet. Apart from it, people can also do financial transactions using Internet banking on their cellular phones or personal digital assistants. Internet banking offers large number of benefits for people involved in financial transactions. There is no need to visit your bank every time you need to transfer money. You can do so by internet banking from the comfort of your home. With net banking facility, one can not only transfer money, but also pay bills, check bank statements, check account balance, request for check book and various other financial transactions. Internet banking has become widely popular among the masses because of its wide array of benefits. All banks offer the online banking facility for their customers nowadays.

Online banking has made the lives easier for people who are too busy to go to bank for conducting their financial transactions. Net banking offers the flexibility to do financial transaction on any day irrespective of the time. In today's fast paced life, people are too much stressed out because of their work pressure and net banking offers them peace of mind as they can pay their bills, book their tickets, do online shopping, etc. by relaxing on couch in their home. Best part of net banking is that it is very easy to do any transaction over the net and highly secure website takes care of all your worries.

Banks have designed their websites in a very user-friendly manner for net banking facilities. Most of the banking interfaces are easily viewable and instructions are provided at every step so that people can carry out any transaction almost effortlessly. The Internet banking facilities varies from bank to bank. One should read the net banking guidelines thoroughly before conducting financial transactions over the internet.

### 3.1. Internet Banking Vital Features

Net Banking has three basic features. They are as follows:

- The banks offer only relevant information about their products and services to the mass.
- Few banks provide interaction facility between the banks and its customers.
- Banks are coming up with arrangements of utility payments, like telephone bills, electricity bills, etc.

### 3.2. Advantages of Internet Banking

An internet banking account is simple to open and use. Internet banking costs less. You can access the information anywhere that you have access to the Internet. It makes your financial life much easier to manage.

*(a) Pay Your Bills Online*

You can use online banking to pay your bills. This will eliminate the need for stamps and protect our self from the check being lost in the mail. Most banks will have a section in which you set up payees. You will need to fill out the information once, and then you can simply choose that profile every time you pay a bill online. If your bank will not pay bills online you may consider paying online through the company.

*(b) View Your Transactions*

Online banking allows you to access your account history and transactions from anywhere. This is the quickest way to check and see if a transaction has cleared your account. This can help you to find out the amount of a transaction after you have lost your receipt. It also allows you to find out about unauthorized transactions more quickly.

*(c) Transfer Money Between Accounts*

Online banking also allows you to transfer money between accounts much more quickly. It is more convenient than using the automated phone service, and can save you a trip to the bank. When you apply or set up your online banking, be sure that all of the accounts you have at the bank are listed. This will make it easier to transfer money and make loan payments online.

*(d) Protect Yourself Online*

It is important to be careful when banking online. You do not want your safety or privacy to be breached. It is important to clear your cookies after each banking session, if you are at a public computer. Additionally you need to make sure that your password is long enough to prevent it from easily being hacked.

## 4. INTERNET BANKING EXISTING SYSTEM

The following Figure: 2 shows three tired architecture, it has three Components, they are

**Client:** There will be two clients for the application. One will be a web-based user-friendly client called bank customers. The other will be for administration purposes.

**Application Server:** It takes care of the server script, takes care of JDBC-ODBC driver, and checks for the ODBC connectivity for mapping to the database in order to fulfill client and administrator's request.

**Database:** Database Servers will stores customer's and bank data.

To use the services provided under Internet banking has following characteristics of Speedy Transactions, Faster processing, No need to stand in Queue, Lower transaction cost, Better utilization of resources, One click Access to Wide-Products. But due to the problem of service disruption of internet banking due to ineffectiveness of the existing system as to the service dis-contentment among the services and service monitoring of the internet banking. We propose Architecture Pattern styles and suggestive framework for effective utilization of Internet banking from the Customer Pattern.

### 4.1. Problem in Existing System

The main issue in the existing architecture "the server cannot able to tackle the over load". The server doesn't know to prioritize the request from numerous clients and so it suffers from overloading. No load the server gets slow. The main problem here is, there is no usage of patterns.

## 5. PROPOSED SYSTEM

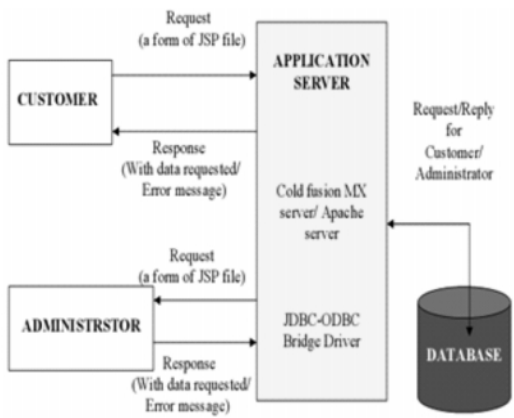

**Figure 2:** Three-tiered Architecture

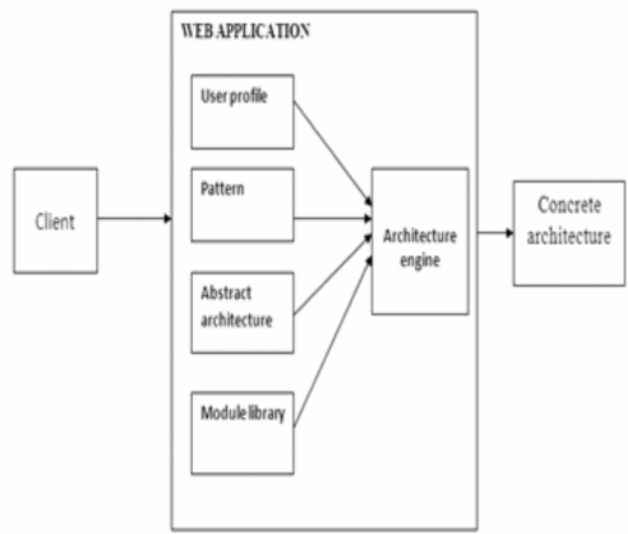

**Figure 3:** Adaptive Architecture for Online Application

Figure: 3 shows user profile, pattern, abstract architecture, module library

### 5.1. User Profile

It is nothing but, based on the age of the client the service will be provided.

| | |
|---|---:|
| Age 18-25 | 35 |
| Age 26-35 | 48 |
| Age 35-45 | 8 |
| Age>45 | 9 |
| Student | 21 |
| Professional | 18 |
| Employed | 52 |

From the figure: 4 consider 18-25 age group clients.

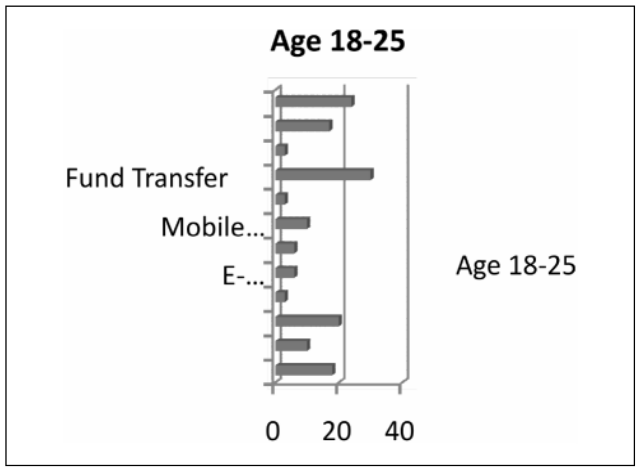

**Figure 4:**　Age 18-25

**Observation**

- When we set the threshold level to 15 the added services to be provided to this user group would be fund transfer, third party services & railway ticket booking. These services would be provided secondary after the fund transfer service.
- Other services can be provided as an added on services for the order of loading for this particular treshold can be as Mobile Recharge, Payment of Bills, Cheque book request, tax payments, third party payments
- When we set the treshold level to 20 the services to be loaded are primarily fund transfer, transaction history would be loaded in the primary slot Railway-ticket or ticket booking, Third-Party transfers would be offered as the secondary services Account statement, transaction history, payment of bills, mobile Recharge, online –DD Payment , cheque book request would be loaded as third services

### 5.2. Pattern

A pattern has been defined as an idea that has been useful in one practical context and will probably be useful in others. Patterns are considered to be a way of putting building blocks into context. For example: to describe a reusable solution to a problem. Building blocks are what you use: patterns can tell you how you use them, when, why and what trade-offs you have to make in doing so.

Patterns offer the promise of helping the architect identify combinations of architectural and/or solution building blocks that have been proven to deliver effective solution in the past, and may provide the basis for effective solutions in the future. Software and building architects have many similar issues to address, and so it as natural for software architects to take interest in patterns as architectural tools.

### 5.3. Abstract Architecture

This architecture is common to all kind of age groups.

### 5.4. Module Library

This block contains the functions and methods based on the service invoked. Based on the user's age corresponding methods and functions will be automatically invoked

### 6. CONCLUSION AND FUTURE WORK

We have examined that, so far patterns are not used in banking application .by using patterns in the internet banking, tremendous benefits can be achieved such as reliability, scalability, performance, availability and load balancing. So the usage of pattern plays a prominent role in online application. While keeping the system at optimum utility, the general layer of service level Framework is functioning, when the users of different category logs into the system & requests for the service, the portfolio of services according to the different age-group is loaded into the server which reduces the load on the system.

The future developments in this framework will be to analyze the patterns emerging out of the utilization of the services in the field of neural networks to be able to cater to different model of customized services to the customers.The future improvements that can be related to the service utilisation and service catering response time of the server and the process improvement steps.